# Frequency-modulated continuous-wave laser distance measurement system using Fabry-Perot cavity as measuring reference

**GUANG SHI,**[1, 2,*] **KEFEI HEI,**[2] **WEN WANG,**[1] **AND NANDINI BHATTACHARYA** [2]

[1] *School of Mechanical Engineering, Hangzhou Dianzi University, No. 1158 the No. 2 Street, Hangzhou, 310018, China*
[2] *Optics Research Group, Department of Applied Sciences, Technical University Delft, Lorentzweg 1, 2628 CJ Delft, The Netherlands*
*\*shiguang@hdu.edu.cn*

**Abstract:** Frequency-modulated continuous-wave (FMCW) is a ranging technique that allows for high precision distance measurement over long distances. Scanning nonlinearity and range of the tunable laser are the main factors affecting the measurement accuracy. Frequency-sampling method is a recognized post-processing scheme to compensate the scanning nonlinearity. In this work, an FMCW laser distance measurement system using a high fineness Fabry-Perot (F-P) cavity as a sampling reference is demonstrated. The frequency of the resampled signal is calculated with a Hilbert transform. The high stability of the F-P cavity and the advantages of the Hilbert transform lead to a high measurement precision when an external cavity diode laser (ECDL) with a scanning range of tens of GHz is available. In this experiment, the scanning range of the ECDL is only 88 GHz, and a measurement uncertainty of 76.8 $\mu$m (with coverage factor of $k = 2$) within a distance of 6.7 m is demonstrated.

## 1. Introduction

Frequency-modulated continuous-wave (FMCW) laser distance measurement also called frequency-scanning interferometry (FSI) is a promising measuring method because of its high accuracy at long ranges. The FMCW interferometry was firstly developed in the radio-frequency region in 1950s [1]. And the same principle has also been used for fiber reflectometry [2]. Early in 1988, an air spaced FMCW laser distance measurement system using a single mode AlGaAs laser diode was reported for short distance ranging [3]. However, due to the limited performance of the laser source, the measurement range and accuracy was low. For an FMCW laser distance measurement system, the measuring range and theoretical measurement accuracy is dependent on the coherence length and the tuning range of the laser source respectively [4, 5]. So once the external cavity diode laser (ECDL) became available as a coherent wavelength tunable source, which can supply a wide tuning range with narrow line width, it became possible for FMCW laser distance measurement to achieve long distance and high-precision [6]. Nowadays, high accuracy absolute distance measurement is an important feature of FMCW laser distance measurement technology, which made it quite valuable for metrology [7], robotic vision [8] and non-contact surface profiling [9].

There are mainly two signal processing methods used in FMCW laser distance measurement technology. The first one is usually called frequency-scanning interferometry (FSI). In this method the measured distance $L$ can be estimated by $L=c·\Delta\Phi/(2\pi·B·n_g)$, where $B$ is the optical frequency scanning range of the laser, $\Delta\Phi$ is the phase change of the interference signal, $c$ is the speed of light in vacuum and $n_g$ is the group refractive index of air [10, 11]. The measurement accuracy of $L$ is dependent on the measurement accuracy of $B$ and $\Delta\Phi$. The major contributor to the final measurement uncertainty is the uncertainty of $\Delta\Phi$. The nonlinearity of the optical frequency scanning and the vibration of the target will seriously affect the measurement accuracy of $\Delta\Phi$. Kalman filter method is proposed to compensated the measurement errors caused by low frequency vibration of the target [12-14] and scanning nonlinearity of the laser [15]. However, for a random vibration of the target or mode hopping of the laser this method would be invalid.

The second signal processing method is called FMCW laser distance measurement. The distance of the target is proportional to the frequency of the interference signal. The distance $L$ can be estimated by $L=c \cdot f/(2\beta \cdot n_g)$, where $f$ is the frequency of the interference signal, $\beta$ is the optical frequency scanning speed of the laser [4]. Here the nonlinearity of the optical frequency scanning is again the main error source. Usually, this can be resolved by the active linearization technique and post-processing schemes [16]. A high-precision and high-stability frequency reference can be used to calibrate the scanning nonlinearity. In 2015, Mateo *et al.* used a molecular frequency reference to calibrate the scanning linearity of the laser [17]. Moreover, the Phase-Locked Loop (PLL) technique is also a feasible solution for the necessity of active linearization. In 2009, Roos *et al.* proposed an active feedback system to maintain the scanning linearity of the tunable laser, in which a fiber-based self-heterodyne interferometer and a frequency standard is employed to detect linear sweeps residuals. And 31$\mu$m range resolution at the distance of 1.5m was demonstrated [18]. In 2016, Xie *et al.* used an ultra-short delay-unbalanced interferometer to suppress the scanning nonlinearity of a tunable laser [7]. In 2017, Behnam *et al.* proposed an FMCW 3D imaging system with an electro-optical phase-locked loop (EO-PLL) for calibrating the sweeping error of the tunable laser. The distance measurement precision was 8 μm for a measurement range of 1.4m [8]. In order to get higher measurement accuracy and measurement speed, a high speed, low noise and complex active feedback system is needed for the active linearization technique.

The frequency-sampling method is a recognized post-processing scheme. In 2011, Baumann *et al.* used a femtosecond optical frequency comb as an optical frequency standard to resample the interference signal at equal frequency steps [19, 20], and demonstrated a precision of 10 *μm* at the distance of 10.5 m. Instead of a high cost optical frequency comb, a fiber Mach–Zehnder Interferometer (MZI) is a more cost effective choice in the frequency-sampling method [21]. Hao Pan *et al.* added a HCN cavity to the FMCW laser distance measurement system and used multiple signal classification (MUSIC) algorithm to enhance the measurement precision to 45 *μm* within 8 m [22]. Cheng Lu *et al.* added a laser Doppler velocimetry (LDV) component to the measurement system to compensate for the environmental vibration. The measurement uncertainty of 8.6 *μm* + 0.16 *μm/m·L* (*k=2*) within a distance range of 1 m to 24 *m* was achieved [23]. In order to satisfy the Nyquist sampling theorem, the optical path difference of the Mach–Zehnder interferometer should be at least twice that of the measuring interferometer. With the increase of measuring range, the length of the optical fiber must be increased accordingly, so that the effects of fiber jitter [24] and dispersion will become more and more significant. Hao Pan [16] and Guodong Liu [25], proposed different algorithms to eliminate the dispersion mismatch.

In this paper, we propose a novel FMCW laser distance measurement with a F-P cavity as the auxiliary interferometer. Compared with a fiber MZI, the F-P cavity is more stable and traceable, moreover, because the F-P cavity is placed in vacuum, the problem of dispersion mismatch does not exist for the distance measurement system. We also propose an unwrapping algorithm, which allows an arbitrary distance to be measured even though the length of the F-P cavity is only 101.768 *mm*. The read-out distance result is calculated using a Hilbert transform. The measurement result is determined by the linear fit of the phase of each resampled point. So the influence of external random noise can be partially eliminated, and the reliability of the result is higher than FSI. More important, for this method, the measurement resolution is not dependent on the formula $\Delta L=c/2n \cdot B$, where $B$ is the scanning range of the tunable laser[4, 18, 20], which is suitable for all FMCW laser distance measurement systems mentioned above. So for this FMCW laser distance measurement system, the requirements for laser scanning range can be reduced. In the experiment, the scanning range of the tunable laser is only 88 GHz. Comparing with a commercial fringe counting interferometer, a measurement uncertainty is 76.8 *μm,* with coverage factor of *k* = 2 has been shown with the measuring range from 2200 mm to 6700 mm.

## 2. Methodology

The length measurement technique we propose and demonstrate in this work is based on interferometry with a frequency modulated external cavity diode laser (ECDL) which uses a Fabry-Perot (F-P) interferometer to sample the interference fringes. This measurement is then compared with a commercial Helium-Neon (HeNe) fringe counting interferometer.

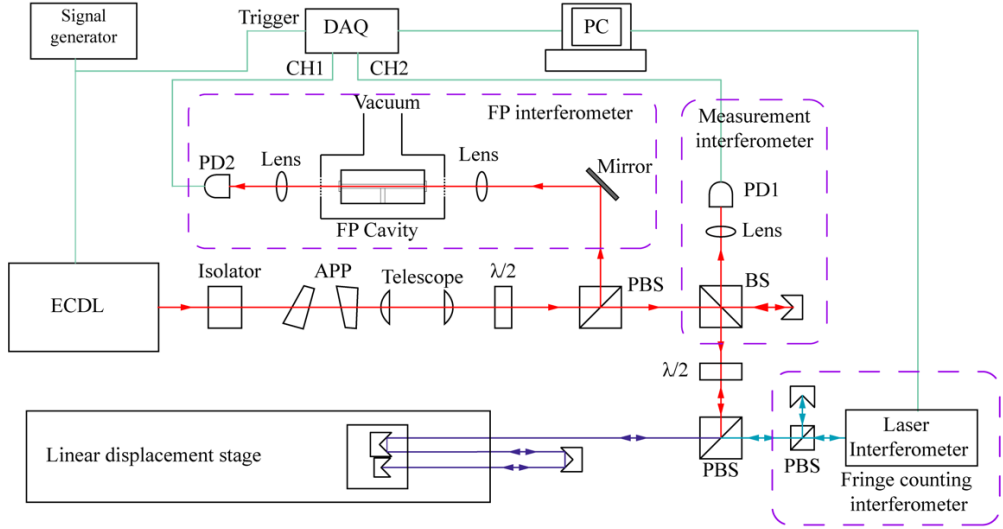

Fig.1. Schematic of the FMCW laser ranging system. ECLD: external cavity laser diode, APP: anamorphic prism pair, BS: beam-splitter, PBS: polarizing beam-splitter, DAQ: data acquisition, PD: photo detector.

Fig.1 illustrates the schematic of the proposed FMCW laser ranging system. This system mainly consists of five parts: a tunable laser source, a Fabry-Perot (F-P) interferometer, a measurement Michelson interferometer, a commercial HeNe fringe counting interferometer and a control & DAQ system. Directly after the ECDL and the optical isolator, a few optical components prepare the beam for use in the rest of the setup. After the initial beam-shaping, the light is divided into two paths using a polarizing beam splitter. The half-wave-plate placed before the beam splitter can be rotated to vary the splitting ratio between the F-P interferometer getting 20% and the measurement interferometer the rest. The measurement interferometer is a typical Michelson with photodetector 1 (PD1) measuring its output, similarly PD2 measures the output of the F-P interferometer.

The beat signal detected by the PD1 at the output from the measurement interferometer can be expressed as:

$$I(\upsilon) = I_0 \cos(2\pi \cdot D \cdot n \cdot \upsilon / c) \tag{1}$$

where $I_0$ denotes the amplitude of the interference signal, $\upsilon$ denotes the instantaneous optical frequency, $D$ denotes the optical path difference of the measurement interferometer, $n$ denotes the phase refractive index of air for a wavelength of $\lambda = c/\upsilon$ and $c$ is the velocity of light in vacuum. Defining the argument of the cosine as

$$\phi = 2\pi \cdot D \cdot n \cdot \upsilon / c, \tag{2}$$

Taking the derivative with respect to frequency we get

$$\frac{d\phi}{d\upsilon} = \frac{2\pi D}{c}(n + \upsilon \frac{dn}{df}) = \frac{2\pi D}{c} n_g, \tag{3}$$

here $n_g$ is the group refractive index defined as [26]

$$n_g = n + \upsilon \frac{dn}{df}. \tag{4}$$

Then the length to be measured $L$ is only determined from the slope $d\phi/d\upsilon$ [27]

$$L = \frac{D}{2} = \frac{d\phi}{d\upsilon} \cdot \frac{c}{4\pi n_g}. \tag{5}$$

Since $I(\upsilon)$ represents a function of instantaneous optical frequency $\upsilon$, if it is resampled with equal optical frequency intervals $\Delta\upsilon$, the measurement signal would become

$$I(p) = I_0 \cos(\frac{4\pi \cdot n_g \cdot L \cdot \Delta\upsilon}{c} \cdot p + \frac{4\pi \cdot n_g \cdot L \cdot \upsilon_0}{c}) \tag{6}$$

where $p = 1, 2, \ldots, N$, and N represents the number of sampling points, $\upsilon_0$ denotes the initial optical frequency.

For the F-P interferometer, the difference between two successive resonance frequencies, the free spectral range (FSR) $\Delta\upsilon_{FSR}$ [28] is given by the equation below.

$$\Delta\upsilon_{FSR} = \frac{c}{2 \cdot n \cdot d_{FP}} \tag{7}$$

where $d_{FP}$ denotes the length of the F-P cavity.

F-P resonance peaks can be applied to sample the interference signal with equal optical frequency intervals the FSR. Then, the measurement interference signal could be further sampled at $\Delta\upsilon_{FSR}$, which is calculated as

$$I(i) = I_0 \cos(\frac{4\pi \cdot n_g \cdot L \cdot \Delta\upsilon_{FSR}}{c} \cdot i + \frac{4\pi \cdot n_g \cdot L \cdot \upsilon_0}{c}) \tag{8}$$

where $\Delta\upsilon_{FSR}$ denotes the FSR of the F-P cavity, $i = 1, 2, \ldots, N$, and N denotes the number of sampling points.

where $d_{FP}$ denotes the length of the F-P cavity.

$$I(i) = I_0 \cos(\frac{2\pi \cdot n_g \cdot L}{d_{FP}} \cdot i + \frac{4\pi \cdot n_g \cdot L \cdot \upsilon_0}{c}) \tag{9}$$

When $n_g \cdot L < 0.5 d_{FP}$, the Nyquist's sampling theorem is satisfied. Then the measured target distance $L$ is proportional to the frequency of the resampled signal $I(i)$.

$$L = f_I \cdot d_{FP} / n_g \tag{10}$$

Where $f_I$ is the frequency of resampled signal $I(i)$.

When $0.5 \cdot d_{FP} < n_g \cdot L < d_{FP}$, the distance of the measured target can be obtained by

$$L = (1 - f_I) \cdot d_{FP} / n_g \tag{11}$$

When the distance of the measured target is larger than $d_{FP}$, $L$ cannot be obtained by Eq.(10) or Eq.(11). Then the absolute distance can be obtained by the following equation.

$$L = \begin{cases} (0.5 \cdot m \cdot d_{FP} + L_t)/n_g & \text{when m is an even number} \\ [0.5 \cdot (m+1) \cdot d_{FP} - L_t]/n_g & \text{when m is an odd number} \end{cases} \tag{12}$$

where $m$ is the integer part of an average value of number of peaks and valleys of the measurement interferometer signal between every two peaks of the F-P interferometer signal in a single measurement and $L_t = f_I \cdot d_{FP}$ is the distance read out.

The uncertainty of the FMCW laser distance measurement system $u_L$ can be obtained from (12)

$$u_L = \begin{cases} \sqrt{(\dfrac{0.5 d_{FP}}{n_g} \cdot u_m)^2 + (\dfrac{0.5m + f_I}{n_g} \cdot u_{d_{FP}})^2 + (\dfrac{d_{FP}}{n_g} \cdot u_{f_I})^2 + (\dfrac{0.5 \cdot m \cdot d_{FP} + f_I \cdot d_{FP}}{n_g^2} \cdot u_{n_g})^2} & \text{when } m \text{ is an even number} \\ \sqrt{(\dfrac{0.5 d_{FP}}{n_g} \cdot u_m)^2 + (\dfrac{0.5m + 0.5 - f_I}{n_g} \cdot u_{d_{FP}})^2 + (\dfrac{d_{FP}}{n_g} \cdot u_{f_I})^2 + \left[\dfrac{0.5 \cdot (m+1) \cdot d_{FP} - f_I \cdot d_{FP}}{n_g^2} \cdot u_{n_g}\right]^2} & \text{when } m \text{ is an odd number} \end{cases} \quad (13)$$

The major contributor to the final uncertainty is the uncertainty of $f_I$ and $d_{FP}$. Because $m$ is an integer, the uncertainty of $m$ is 0. In our experiment, the central wavelength of the ECDL and the wavelength of the HeNe fringe counting interferometer are both 633 nm, the uncertainty of $n_g$ almost does not affect the results. So Eq.(13) can be simplified to

$$u_L = \begin{cases} \sqrt{(\dfrac{0.5m + f_I}{n_g} \cdot u_{d_{FP}})^2 + (\dfrac{d_{FP}}{n_g} \cdot u_{f_I})^2} & \text{when } m \text{ is an even number} \\ \sqrt{(\dfrac{0.5m + 0.5 - f_I}{n_g} \cdot u_{d_{FP}})^2 + (\dfrac{d_{FP}}{n_g} \cdot u_{f_I})^2} & \text{when } m \text{ is an odd number} \end{cases} \quad (14)$$

In order to obtain the frequency of the resampled signal $I(i)$, fast Fourier Transform (FFT) is the most direct and convenient method. For a relatively long single frequency signal, a high-resolution frequency measurement result can be obtained from the FFT algorithm. However, for a series of signal with only less than 100 points, the relative resolution is only 1/100, so the resolution of FFT depends on the length of the signal. In our FMCW laser ranging system, the scanning range of the ECDL being only 88 GHz the number of resonances from the F-P cavity overlapping with the measured signal is only 58. Thus the length of the resampled signal is only 58 points. Fig.2 shows the interferometer signal of the measurement interferometer and the F-P interferometer. Fig.3 shows the resampled signal, and the spectrum of the read distance is shown in Fig.4. It can be seen from the figure that the resolution is only 3.5 millimeters.

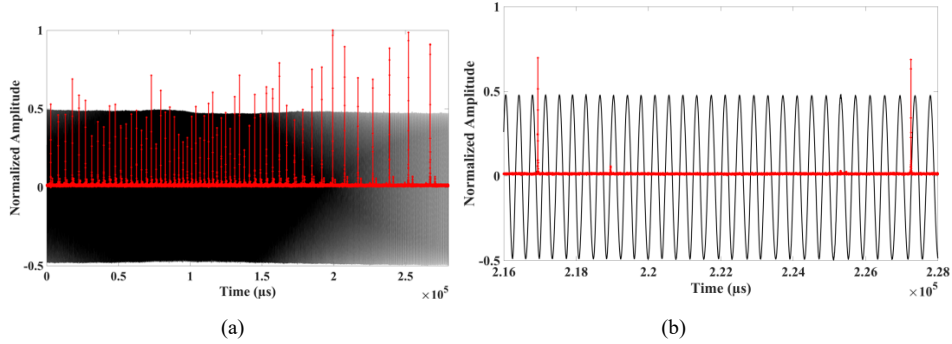

Fig.2. Interference signal of measurement interferometer (black) and F-P interferometer (red). (a) the signal of once measurement. (b) enlarged view of (a)

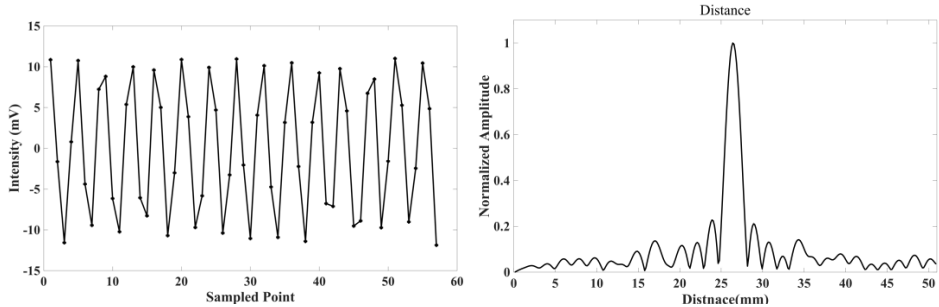

Fig.3. The resampled signal          Fig.4. The distance spectrum

Hilbert transform is another method to calculate the frequency of single frequency signal. As is shown in Fig. 5, we performed Hilbert transform on the resampled signal to obtain the instantaneous phase. The unwrapped instantaneous phase curve is shown in Fig.6. The resolved phase is shown in black, and the result of least square fitting of the phase is in red. The slope of the fitted line is 1.625±0.004, with 95% confidence bounds. The frequency of the signal shown in Fig5 is calculated to be $f_1$ =0.2586±0.00064, again with 95% confidence bounds, and the corresponding distance range is estimated to be 26.314±0.064 mm.

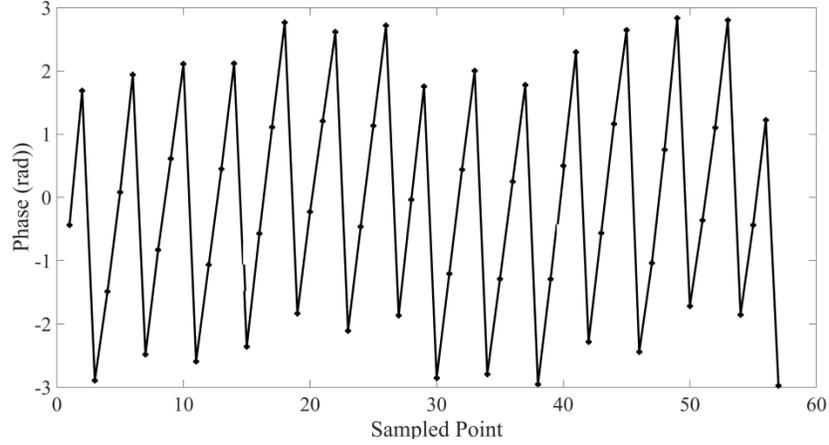

Fig.5. The phase obtained from the Hilbert transform of the resampled signal shown in Fig.3.

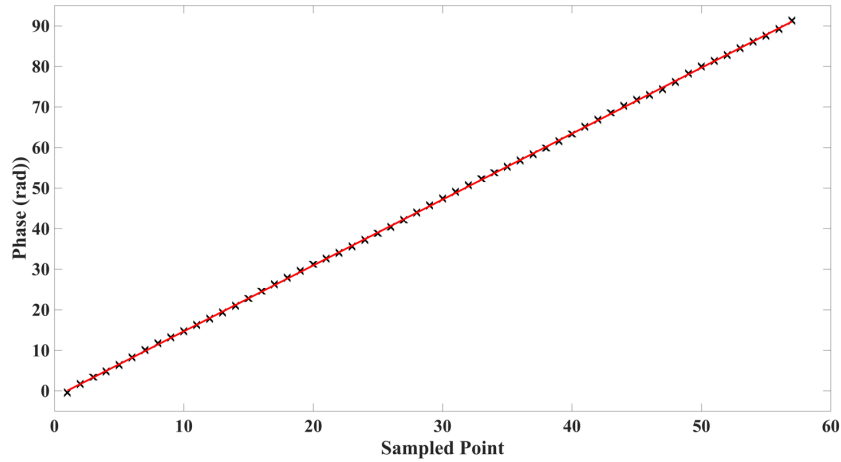

Fig.6 The phase after unwrapping (black) and is then fitted by the least squares function (red).

### 3. Experiment

Fig.1 illustrates the schematic diagram of the actual experimental FCWM laser ranging system. In this system, an ECDL (Newport TLB-6210) is chosen as the tunable laser source with the central wavelength of 633nm, and the scanning range available is 88GHz. The anamorphic prism pair and the telescope are employed to change the elliptical beam cross-section of the laser diode into a circular one, and to improve the collimation of the measurement beam. The signal generator is employed to supply the scanning signal at the frequency of 2 Hz. The Fabry-Perot cavity (custom built, Research Electro-Optics, finesse is 11041) is made out of ultra-low expansion (ULE) glass with length of 101.768 mm and is placed in vacuum.

The Michelson measurement interferometer is an unbalanced interferometer built with 1 inch sized optics, using a non-polarizing beam splitter (BS) and retroreflector prisms, whose long arm is used as the length to be measured. The measurement arm has a maximum length of 1.5 m and consists of a long rail with electric carriage carrying two retroreflector prisms. As is shown in Fig.1, The measured optical path is folded three times by 3 retroreflector prisms. At the nearest measurement position of the rail, the distance difference of the two arms of the measurement interferometer is 2208.877 mm. And this position is set to be measurement origin of the fringe counting interferometer. The commercial HeNe fringe counting interferometer (Agilent 5519A), with a linear distance measurement accuracy of ± 0.4 ppm in air is used for the comparison measurement.

The beam from the ECLD and the counting interferometer is combined with a polarizing beam-splitter (PBS). The auxiliary mirrors in the optical path are adjusted to ensure that the two beams overlap optimally. As a result, both beams largely propagate through the same volume of air, having a shared measurement arm. Then the half-wave-plate is employed to separate the returning beams after the PBS. The light from the HeNe laser propagates back through the same path to the HeNe laser head which contains the detector for fringe counting. The interferometer signal of the Michelson measurement interferometer is detected by PD1. The detected signal of PD1 and PD2 are synchronously sampled by a DAQ board, with a sampling rate of 1MHz. During the experiment, environmental conditions were measured by a VAISALA air parameter sensor. The temperature, humidity and atmospheric pressure were measured to be 27±0.2 ℃，46.2±0.2% and 101.61±0.02 KPa.

### 4. Results and discussion

In the experiment, the carriage is physically moved from 0 mm to 1500 mm, which corresponds to an optical path length of 0 mm to 4500 mm, with 10 measurements obtained at each position. Moreover, the displacement of the carriage is measured by both FMCW laser distance measurement system and the counting interferometer simultaneously. Fig. 7 shows the differences between the individual measurements done by FMCW laser ranging system and the distance measured by the HeNe fringe counting interferometer. The central wavelength of the ECDL and the wavelength of the fringe counting interferometer are both 633 nm so the refractive index is assumed to be to the same in their ranging formulas. The zero position is based on the average of 10 measurements. For each individual measurement, the agreement between the FMCW ranging system and the fringe counting interferometer is within 100 μm. When averaged over ten measurements, the largest difference is 28 μm. The standard deviation does not show a clear distance dependence and is on average 34 μm. Because environmental effects such as turbulence and vibrations will affect a single measurement result, averaging multiple measurements can improve the measurement accuracy.

The F-P cavity is located inside a small vacuum vessel to prevent any changes in air pressure and temperature from influencing the FSR. In the experiment, uncertainty of temperature is 0.2 ℃. The thermal coefficient of ULE is smaller than ±30·10$^{-9}$ /K. The fluctuations of the temperature of the cavity itself will be much lower because of its large thermal mass. So the effect of temperature on FSR is small enough to be ignored. The uncertainty of air pressure in the small vacuum vessel is 0.05 KPa. Substituting temperature and humidity parameters, calculate by using Edlen's equation, the uncertainty of the refractive index in the F-P cavity is $u_n$ = 0.00000013, the uncertainty of FSR is $u_{FSR}$ = 180 Hz. Assuming $n_g$=1, from Eq(8) the uncertainty of physical length of the cavity (with coverage factor of k = 2) is $u_{dFP}$ = 0.00013 mm.

In the experiment, the absolute distances obtained from the FMCW laser ranging system vary from about 2200 mm to 6700 mm. The range of $m$ in eq (12) and (14) is from 43 to 131. The measurement uncertainty of the resampled frequency is $u_{fl}$ = 0.0007 (with coverage factor of $k$=2). At each position, substituting $n_g$=1, $m$= 43~131, $u_{dFP}$ = 0.00013 mm and $u_{fl}$ = 0.0007 into the Eq.(14) gives $u_L$=70~70.5 μm (with coverage factor of $k$ = 2). Because the value of $m$ is not large, the contribution to measurement uncertainty of $u_{dFP}$ is small. The main uncertainty contribution is $u_{fl}$. From fig.7, it can be calculated that the standard deviation the measurement is $\sigma = 34$ μm, and the Type A estimate of uncertainty (with coverage factor of $k = 2$) is $u_A = 2.26\sigma = 76.8$ μm, which matches the calculation result. Path length changes due to vibrations, uncertainty in refractive index of air, the inaccurate of resampling position and the stability of the intensity of the ECDL are the main reasons result in the measuring uncertainty of $f_l$. If the scanning range of the ECDL was enlarged, more resample points could be obtained and smaller measurement uncertainty could be obtained.

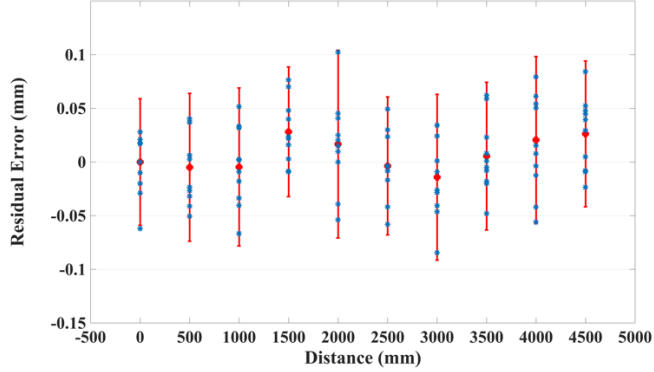

Fig.7. The Measured differences between distance measurement with the FMCW laser distance measurement system and a HeNe fringe counting interferometer. The red bar denotes twice of the standard deviation of the measurements.

As mentioned above, arbitrary distances were chosen, but path length differences very close to $L_d = 0$ or $L_d = d_{FP}/4$ were avoided. At $L_d = 0$ m all wavelengths have the same phase (neglecting nonlinear air dispersion), so a typical cosine dependence like in Fig. 3 is not observed. Close to $L_d = d_{FP}/4$ the Nyquist frequency is approached and each period of the cosine is only determined by 2 points. In order to overcome the issue, a multiplex scheme could be envisaged. We can add another reference path to the measurement interferometer. By using a beam splitter and two shutters, then the reference path can be always selected such that the path length differences close to $L_d = 0$ m or $L_d = d_{FP}/4$ do not occur.

## 5. Conclusion

In this paper, we demonstrate an FMCW laser distance measurement system with F-P interferometer as measuring reference. An absolute distance measurement at the range between 2200mm and 6700 mm has been completed. For tradition FMCW laser distance measurement system, the measurement resolution depends on scanning range of the laser source ($\Delta = c/2B$ where $B$ is the scanning range). The scanning range of the ECDL in this experiment is only 88 GHz, the corresponding measurement resolution is only 3.4 mm. However, for our distance measurement system, in comparison to a HeNe fringe counting interferometer, a measurement uncertainty of 76.8 μm (with coverage factor of $k = 2$) has been shown. This method employs an F-P cavity as the measurement reference instead of a long fiber. The F-P cavity is placed in vacuum making the optical length much more stable than the conventionally used long fiber, also dispersion mismatch is avoided in this scheme. Using the Hilbert transform algorithm and least square method to calculate the frequency of the resampled signal leads to an averaging effect in the analysis. This decreases the influence of environmental effects such as turbulence and vibrations on a single measurement result. Thus the reliability of a single measurement is better than FSI. In general, this FMCW laser distance measurement system reduces the requirement for laser scanning range and eliminates the dispersion mismatch. Another advantage is the ability to resist environmental disturbances better than fiber reference FMCW distance measurement systems and FSI systems.

**Funding**

This work was supported by the National Natural Science Foundation of China under Grant No. 51505113; Zhejiang Provincial Natural Science Foundation of China under Grant No. LQ16E050002, and Grant No. LZ16E050001; and State Key Laboratory of Precision Measuring Technology and Instruments Project under Grant No. PIL1601.